\documentclass[%
aip,
 amsmath,amssymb,
 preprint,%
 reprint,%
]{revtex4-1}

\usepackage{graphicx}
\usepackage{dcolumn}
\usepackage{bm}
\usepackage{xcolor}
\usepackage{booktabs}
\usepackage{float}

\usepackage[utf8]{inputenc}
\usepackage[T1]{fontenc}
\usepackage{mathptmx}
\usepackage{lineno}
\usepackage{soul}
\usepackage{hyperref}
\usepackage{siunitx}

\newcommand{\man}{\texttt{Manager} }
\newcommand{\worker}{\texttt{Worker} }
\newcommand{\workers}{\texttt{Workers} }

\begin{document}

\title[Parallelized Real-time Physics Codes for Plasma Control on DIII-D]{Parallelized Real-time Physics Codes for Plasma Control on DIII-D}

\author{A. Rothstein}\email{arothstein@princeton.edu}
\affiliation{Princeton University, Princeton, NJ, USA}
\author{K. Erickson}
\affiliation{Princeton Plasma Physics Laboratory, Princeton, NJ, USA}
\author{R. Conlin}
\affiliation{Princeton University, Princeton, NJ, USA}
\affiliation{University of Maryland, College Park, MD, USA}
\author{A. Bortolon}
\affiliation{Princeton Plasma Physics Laboratory, Princeton, NJ, USA}
\author{E. Kolemen}
\affiliation{Princeton University, Princeton, NJ, USA}
\affiliation{Princeton Plasma Physics Laboratory, Princeton, NJ, USA}

\begin{abstract}
    A real-time safe multi-threading library was developed on the DIII-D plasma control system to optimize the real-time TORBEAM and real-time STRIDE physics codes. These physics codes are crucial for future fusion power plant operation as they provide information about electron cyclotron wave propagation and heating as well as inform about ideal plasma stability limits. The real-time TORBEAM code executed consistently in under \SI{20}{ms} while the real-time STRIDE code computes in \SI{100}{ms}. The multi-threading library developed in this work can be applied to other real-time physics-based codes that will be crucial for the next generation of fusion devices. 
\end{abstract}
\maketitle

\section{Introduction}\label{sec:intro}
Tokamaks are currently the most promising path to fusion energy\cite{buttery_advanced_2021}, with future reactors such as ITER\cite{hassanein_potential_2021} and SPARC\cite{creely_overview_2020} promising net positive energy production. Due to the dynamic nature of a tokamak plasma shot, real-time plasma control systems (PCS) are necessary to maintain confinement and have been the subject of study across many tokamak devices\cite{lennholm_plasma_2000,kudlacek_overview_2024,yan_custom_2021,yonekawa_current_2004,margo_current_2020,galperti_overview_2024,de_vries_strategy_2024,perek_preliminary_2025,creely_overview_2020,creely_sparc_2023,hahn_advances_2020}.  

With the extreme robustness needed for a fusion pilot plant PCS, significant research has been done on developing real-time physics-based codes useful for plasma control. These include more standard equilibrium reconstruction codes like real-time EFIT\cite{ferron_real_1998} and LIUQE\cite{moret_tokamak_2015} to more advanced real-time profile prediction codes in RAPTOR\cite{piron_development_2021,felici_real-time_2011} and real-time neutron rate calculation codes\cite{weiland_rabbit_2018}. Additional work has been done to adapt ray tracing codes such as TORBEAM\cite{poli_torbeam_2018} for electron cyclotron heating (ECH) as well as the magnetohydrodynamic (MHD) stability code STRIDE\cite{glasser_robust_2018,glasser_direct_2016,glasser_riccati_2018,glasser_ideal_2020} that can give important information about ideal stability limits. However, while the real-time implementations of these physics codes demonstrate significant latency and throughput performance improvements relative to their offline counterparts, they need well-designed PCS environments to achieve the necessary execution times to be useful for real-time control. 

Single core performance bottlenecks present a recurring need in real-time physics codes is for a robust multi-threading library to enable parallelization across CPU cores. In the case of real-time TORBEAM (rt-TORBEAM), the ray trace calculations are independent across gyrotrons, yielding conceptually simple parallelization. While  existing multi-threading libraries exist such as the widely used OpenMP\cite{openmp_2008}, as discussed later in \autoref{sec:environment}, these do not target systems requiring deterministic latency, such as would be required for any commercial fusion reactor, and are thus not desirable. This motivates the work done here in developing and implementing a real-time-safe multi-threading library on the DIII-D PCS to run both the TORBEAM and STRIDE codes in real-time. 

The rest of the paper is organized as follows: in \autoref{sec:library} we discuss the constraints of the DIII-D PCS environment and the implementation details of the real-time multi-threading library. In \autoref{sec:TORBEAM} we discuss the implementation of the rt-TORBEAM code on DIII-D utilizing the multi-threading library. In \autoref{sec:STRIDE} we explain the STRIDE real-time problem formulation and implementation as well as show experimental results before concluding in \autoref{sec:conclusion}.

\section{Real-time Multi-threading Library}\label{sec:library}

In this section, we first discuss the environment of the DIII-D PCS, then move towards regulations that must exist on a resilient, fusion power plant quality PCS, and finally describe the design of the real-time compatible multi-threading library.

\subsection{PCS Environment}\label{sec:environment}

The DIII-D PCS functions with a fixed number of processes that each run on separate CPU cores. In each process, a number of algorithms execute serially within an allotted cycle time. If they cannot complete before this time elapses, undefined behavior can occur and can lead to cascading errors for algorithms across the PCS. Thus, it is critical that individual algorithms have well-defined behavior and deterministic runtime.

Additionally, for production-grade software that will be run on a fusion reactor, the PCS code must be robust to any possible error in a way that there will be no undefined behavior that could crash the overall PCS and lead to a critical failure. For a fusion reactor on the grid, this would mean an unexpected loss in power production or, in a worst case scenario, a full PCS failure could lead to an uncontrolled disruption of the plasma and damage hardware components of the reactor. To avoid this, the PCS code must be clear about how it will execute in the event of any and all edge cases

The OpenMP library\cite{openmp_2008} is an industry standard multi-threading library with roots in the world of high performance computing.  While it is excellent for most applications, a robust PCS system is not a safe place to apply this package. Many details of thread creation and work assignment are implementation defined and deferred to the respective compilers, and the OpenMP specification does not give hard real-time guarantees required for microsecond scale plasma control. There are no known implementations that provide these details either as implementation defined scope. There are various complicated mechanisms that can mitigate some timing issues and reduce latency concerns (thread pinning, lock elision, etc.), but these are all best effort, subject to change across compiler versions, and uniquely configured between compiler implementations. 

The logical next possibility to consider would be utilizing GPUs connected to the main PCS CPUs for computationally intensive physics codes. While this is an enticing option, GPUs require significantly higher start-up costs than CPU multi-threading and only make sense when there are hundreds or thousands of parallel computations to be done. As described later in \autoref{fig:rt-TORBEAM} and \autoref{fig:rtstride}, these physics codes need further development work to benefit from GPU hardware. While the physics codes described in this text would not benefit from GPU processing, a prime example for utilizing GPUs in the PCS would be machine learning models\cite{conlin_keras2c_2021,rothstein_pacman_2025,boyer_toward_2021,rothstein_torbeamnn_2025,seo_avoiding_2024,anirudh_2022_2023,churchill_deep_2020} that require large matrix operations and can best leverage GPU hardware. 

For the code described in \autoref{sec:TORBEAM}, the calculations were completed on a system with dual 12-core Intel Xeon Gold 6136 CPUs running at 3.0 GHz. For the code described in \autoref{sec:STRIDE}, the calculations were run on a system with quad 18-core Intel Xeon Gold 6154 CPUs also running at 3.0 GHz. Both of these systems isolated the cores in use for real-time from other processes. The former system used a stock CentOS 7 environment with limitations on isolation possibilities, while the latter system used a commercial real-time OS from Concurrent-RT called RedHawk. This OS provides complete core isolation, including from the kernel timer interrupt.  This reduces latency jitter to the minimum allowed by the hardware.

\subsection{Real-time safe multi-threading library details}\label{sec:implementation}

\begin{figure*}
    \centering
    \includegraphics[width=\linewidth]{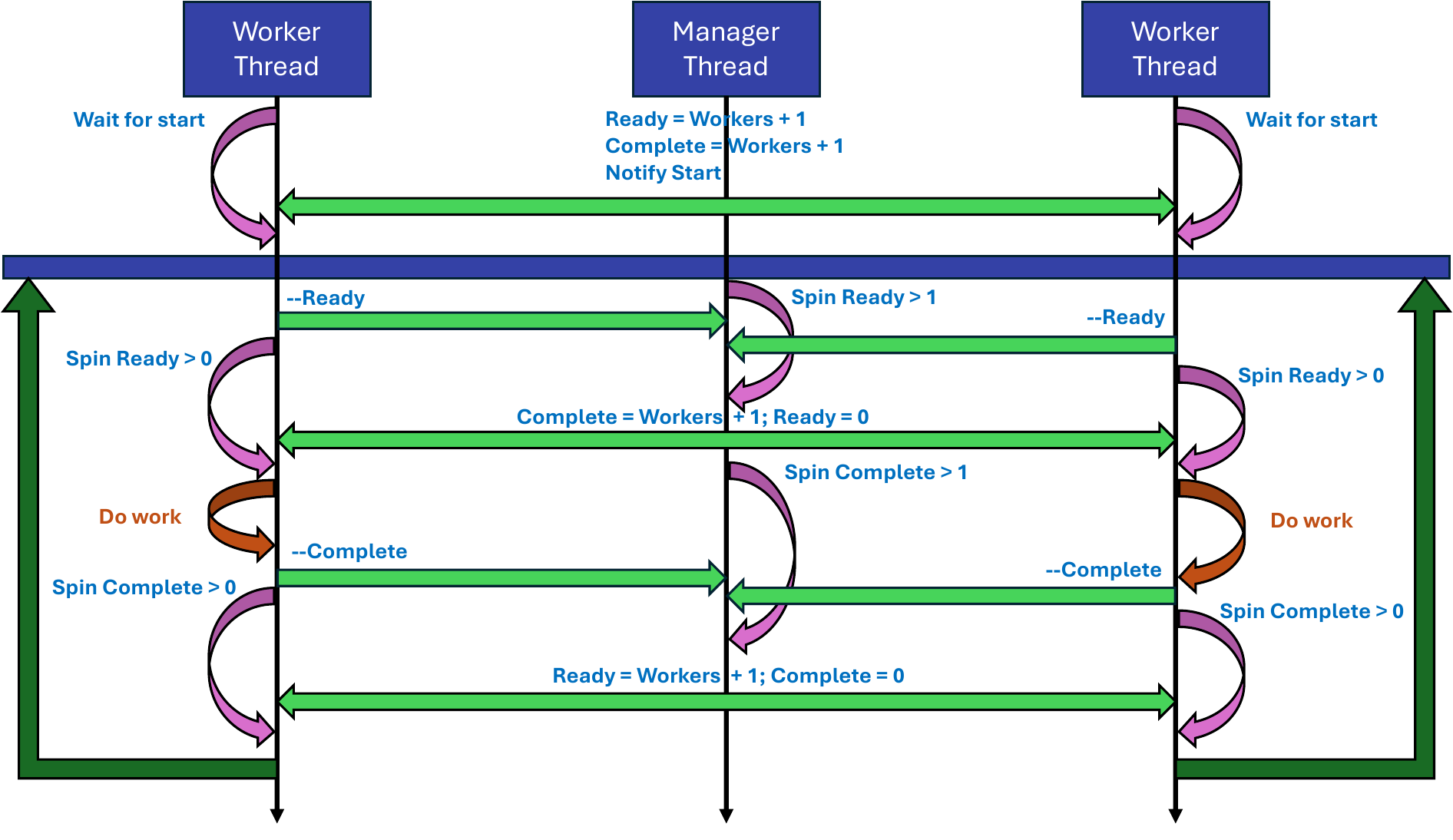}
    \caption{UML diagram of the multi-threading algorithm. The \man and \workers each follow the vertical black lines downward to progress through the shot. The horizontal bright green arrows represent communications between the \man and \texttt{Workers}. An arched pink arrows denotes a thread waiting until some condition \texttt{X} is made \texttt{false} where the condition is given by \texttt{Spin X}. The orange arched arrows denote \workers doing computational work. Finally, the dark green arrows represent the \workers returning to the post-initialization state to repeat for the next CPU cycle during a shot. }
    \label{fig:UML}
\end{figure*}

The multi-threading library described here was initially developed for the Digital Coil Protection System on NSTX-U using a custom \texttt{C++11} environment and known as the Bidirectional Atomic Synchronization System \cite{erickson_nstx-u_2014}. For the DIII-D PCS, that system was adapted to \texttt{C11} and redesigned for more flexible operation while still maintaining the original abstraction goals. Notably, the various components of the threading system are unaware of the infrastructure of the calling context.  While the \texttt{C++} implementation used object oriented based callable function objects with type bound arguments, the \texttt{C} implementation resorts to pthread-style function pointers with fixed opaque pointers to contain any desired payloads.

The library works by having a single Manager Thread (the \man) which assigns work units to individual Worker Threads (the \texttt{Workers}). This is visualized in \autoref{fig:UML} with two \texttt{Workers}. Note the only limit to the \workers count is the available threads from the hardware being used to run the algorithm. In our utilization of this library in the DIII-D PCS, the \man is the same as the main thread for the PCS-controlled CPU. This design decision was made to simplify coordination of background work with the data flow of the PCS algorithm without introducing additional complexities such as futures and promises. The tradeoff for that simplicity is that the PCS CPU blocks waiting for the workers to complete their tasks.

The UML diagram in \autoref{fig:UML} is split into two stages: the pre-shot, non-real-time initialization stage above the blue horizontal block, followed by the during-shot real-time calculation stage below the blue horizontal block.  The during-shot stage repeats until the plasma shot is complete.  Forward progression of the algorithm is represented by the black arrows pointing downwards and the repetition of CPU cycles throughout the shot are given by the dark green arrows on the far left and far right that return to the blue horizontal block. The following two subsections describe the algorithm represented in \autoref{fig:UML}.

All communication in this design uses low level atomic variables with acquire-release memory model semantics as specified by the \texttt{C11} standard. These directly mirror the original design's use of the \texttt{C++11} memory model for the same purpose. The important guarantee from these atomic operations based on the chosen memory model is that when a particular atomic operation becomes visible in another thread, all previous non-atomic operations are also visible. This model does not guarantee ordering of those non-atomic operations, but ordering for this design is not needed. Critically, this approach avoids higher level abstractions that would otherwise involve the kernel, such as semaphores. While those are conceptually simpler and easier to implement, they unfortunately introduce unacceptable nondeterministic behavior from the kernel.

Of particular importance for this mechanism to work is the bidirectional component of the original implementation. Both the \man and the \workers independently notify each other of their status through these atomics, allowing synchronization points at the beginning and end of the critical section. This is analogous, and in fact was modeled after, the Ada language "rendezvous" core feature. Without a rendezvous in both directions, the synchronization would break. This means that threads do spend time spinning idly waiting for other threads to catch up. Balancing workloads to maximize CPU utilization is the job of the developer. Such is the nature of real-time systems, where maximal throughput is secondary while determinism and latency are primary.

\subsubsection{Pre-Shot Initialization}
During the plasma shot set-up, the \man initializes the \workers and each of the \workers waits for the plasma shot to begin. At this stage, the \man enters the state of completed work where each of the \workers has no work (equivalent to having completed all previous work) and is ready to receive new jobs. This is the same state that will be achieved at the end of each CPU cycle during the shot. This begins by setting the variables \texttt{Ready} and \texttt{Complete} to the total number of \workers plus one. Finally, when the shot is about to begin, the \man communicates to each worker that the shot is ready to begin, represented by the green arrow that travels from the \man to the \texttt{Workers}. 

\subsubsection{During Shot Calculation}
Entering the CPU cycle during a shot after the blue horizontal block, the \man has previously notified the \workers that the shot has begun, or all previous work has been completed in the previous cycle. The \man waits for the workers to respond by waiting for \texttt{Ready>1} to become \texttt{false}. At this point, each \worker notifies the \man they are ready by reducing \texttt{Ready} by $1$, as indicated by the horizontal green lines directed from \workers to the \man, and will spin until \texttt{Ready>0} is \texttt{false}.

Once all of the \workers have communicated readiness to the \man, the \man sets \texttt{Ready} to $0$ and distributes the work to be calculated, such as the rt-TORBEAM or rt-STRIDE calculation, as seen in the bright green arrow that is directed from the \man to the \texttt{Workers}. The \man also sets \texttt{Complete} to the total number of \workers plus $1$ and waits for \texttt{Complete>1} to be \texttt{false}. Now, each \worker receives their work from the \man, observes \texttt{Ready=0} and begins their work. When a \worker completes their work, they communicate back to the \man to decrement \texttt{Complete} by $1$ and each \worker waits for the remaining \workers to complete their work, which is when  \texttt{Complete>0} is \texttt{false}. 

Finally at the end of the cycle, the \man gathers the results of the calculations and prepares the \workers for the next CPU cycle. When \texttt{Complete>1} becomes \texttt{false}, all workers have completed their work and the \man sets \texttt{Complete} to $0$ to indicate all \workers have finished and sets \texttt{Ready} to the number of \workers plus $1$ to reset the state for the start of the next cycle. At this point, the \man can do any additional calculations required by the algorithm or communicate results with other PCS algorithms that want the results of the parallelized calculation. 

A key point is that the \workers end up in the same state at the end of a CPU cycle as they were after the pre-shot initialization. This guarantees robust thread management and no undefined or unexpected behavior of individual threads.

\section{TORBEAM Ray Tracing}\label{sec:TORBEAM}
Real-time ECH ray tracing is seen as a requirement for accurate ECH aiming to deposit electron cyclotron current drive (ECCD) at rational surfaces for active NTM control\cite{kolemen_real-time_2013,zohm_control_2007,sauter_requirements_2010,park_initial_2024} as well as a variety of other tasks utilizing ECH and ECCD such as sawtooth control\cite{felici_integrated_2012,goodman_sawtooth_2011,maraschek_active_2005,lennholm_closed_2009,lennholm_real-time_2016}, impurity screening\cite{dux_influence_2003,odstrcil_dependence_2020}, or ELM control\cite{logan_access_2024,hu_effects_2024,lennholm_real-time_2016}. Combined with a basic feedback controller, rt-TORBEAM can be used to accurately steer ECH rays to desired locations in the plasma. This section describes the implementation of the rt-TORBEAM code in  \autoref{sec:rtTORBEAM} followed by experimental deployment in  \autoref{sec:TORBEAMexp}.

\subsection{Problem formulation and real-time implementation}\label{sec:rtTORBEAM}

\begin{figure}
    \centering
    \includegraphics[width=0.8\linewidth]{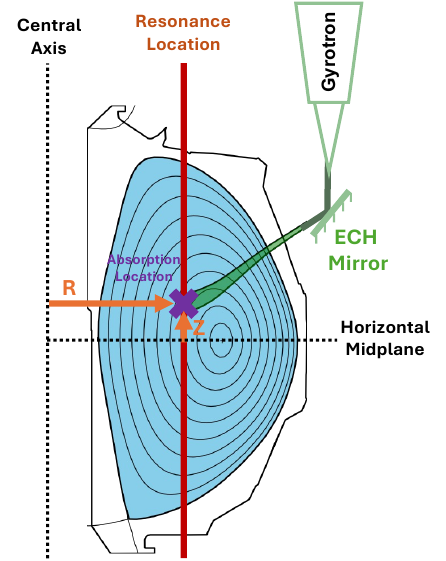}
    \caption{Example ECH ray-tracing where ECH beam begins from the gyrotron, makes its way to the ECH mirror to be launched into the plasma at some poloidal and toroidal angle. It then travels through the plasma at a calculated trajectory until being absorbed at the vertical red resonance location. The maximum absorption location (purple X), is given by coordinates $(R,Z)$. Refer to \citet{poli_torbeam_2018} for further explanation. }
    \label{fig:ECH}
\end{figure}

Each gyrotron on DIII-D has an independent, steerable mirror that aims the ECH ray into the plasma. The rt-TORBEAM code calculates its trajectory until the location of maximum absorption is found or the ray leaves the plasma boundary, the last closed flux surface, as visualized in \autoref{fig:ECH}.  Since each gyrotron launchers and ray trace is fully independent, rt-TORBEAM can be run for all gyrotron launchers in parallel to use the multi-threading library. 

Note that a main difference between the offline and rt-TORBEAM code is the simplification from a Gaussian beam to a single ray. This means the full ECH and ECCD deposition profiles are not calculated by rt-TORBEAM, only the $(R,Z)$ location of maximum absorption is calculated to properly steer the ECH mirror. One option to overcome this would be to run rt-TORBEAM multiple times to simulate a Gaussian beam. Another option, and perhaps more straightforward, would be to utilize machine learning surrogate models to calculate the ECH and ECCD profiles directly\cite{rothstein_torbeamnn_2025}. 

For our application of ECH mirror steering, the user provides a spatial coordinate $\rho$ location, $\rho_{target}$, and the standard rt-TORBEAM library calculates the $(R,Z)$ location of maximum absorption which is converted to a $\rho_{torbeam}$ value based from the real-time equilibria\cite{ferron_real_1998}. $\rho$ is defined as $\rho=\sqrt{\Phi_N}$ where $\Phi_N$ is the normalized toroidal flux to $0$ in the core and $1$ at the last closed flux surface. The difference in $\rho_{target}-\rho_{torbeam}$ is used to change the mirror angle so that the $\rho$ value calculated from rt-TORBEAM matches the desired $\rho_{target}$ value. This steering algorithm was originally implemented in \citet{kolemen_real-time_2013}. 

\subsection{Experimental results}\label{sec:TORBEAMexp}
\begin{figure}
    \centering
    \includegraphics[width=\linewidth]{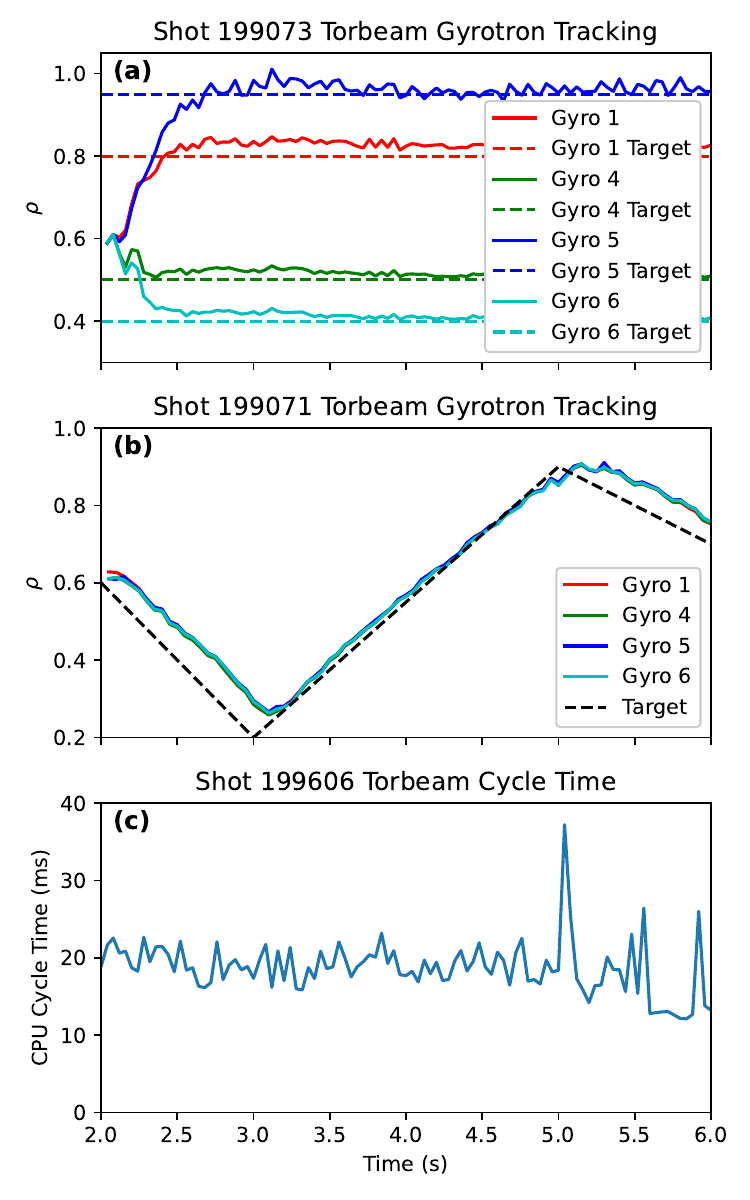}
    \caption{Experimental results of rt-TORBEAM calculation and gyrotron mirror control. \textbf{(a)} DIII-D shot 199073 with constant $\rho_{target}$ values for four controlled gyrotron mirror where for each individual gyrotron, $\rho_{target}$     (dashed line) and $\rho_{torbeam}$ (solid line) have matching colors. \textbf{(b)} DIII-D shot 199071 with dynamic $\rho_{target}$ (black dashed line) that is the same across all gyrotrons (solid colored lines). \textbf{(c)} CPU cycle time to compute rt-TORBEAM for all four gyrotrons. }
    \label{fig:rt-TORBEAM}
\end{figure}

The experimental results from the rt-TORBEAM deployment are shown in \autoref{fig:rt-TORBEAM}. In these results, we only show the results from the rt-TORBEAM calculation as the real-time code has been adequately validated in \citet{poli_torbeam_2018}. These results show $\rho$ tracking for a single mirror and similar results are seen across all steered gyrotron mirrors. The results show reasonable $\rho$ tracking to within $|\rho_{target}-\rho_{torbeam}|\leq0.05$. 

\autoref{fig:rt-TORBEAM}.a shows constant $\rho_{target}$ values for each of the four controlled gyrotrons and the achieved $\rho_{torbeam}$ for each of them. The gyrotron mirrors are slow to move and required approximately \SI{0.5}{s} to move from their initial position at $\rho=0.6$ to their respective $\rho_{target}$ values and would require redesigning the mirror steering hardware to further speed up. There is also some minor steady-state error which is attributed to the differences in real-time equilibria used for control and the offline reproduction of the real-time equilibria used for the conversion from $(R,Z)$ to $\rho$. Finally, it is worth noting that there is more fluctuation in the $\rho=0.95$ target value close to the plasma edge. Near the plasma edge, the ECH deposition is more effected by density fluctuations and a faster real-time calculation and faster mirror hardware would be required to have improved $\rho$ tracking. 

\autoref{fig:rt-TORBEAM}.b shows a triangle waveform for $\rho_{target}$ used for all four gyrotrons. While tracking is good during the center when it initially matches $\rho_{target}$ at \SI{3.2}{s}, it is clear the mirror were too slow to follow the desired $\rho_{target}$ in the two downward slope sections. A gentler slope for $\rho_{target}$ would be needed for improved tracking, or an improved model-based control scheme would be needed to reduce the lag in $\rho$ tracking. Additionally, it is clear at \SI{5}{s} that the mirror struggles to change directions quickly so sharp changes in $\rho_{target}$ seen here are not desirable for the current controller.   

The cycle time plot in \autoref{fig:rt-TORBEAM}.c shows continuous computation in around \SI{20}{ms} for the $4$ gyrotron mirror that were steered in this experiment. Due to the parallel nature of adding additional gyrotrons and utilization of the multi-threading library, the \SI{20}{ms} cycle time will stay constant as DIII-D expands gyrotrons as long as there is a free worker thread for every steered gyrotron mirror. At \SI{5}{s} there is a large spike in the CPU cycle time where it jumps from the consistent \SI{20}{ms} up to nearly \SI{40}{ms}. This large spike is attributed to the rt-TORBEAM code rather than the multi-threading library and further improvements to the rt-TORBEAM code should be able to limit the effect of edge cases greatly increasing execution times. 

\section{STRIDE Ideal Stability}\label{sec:STRIDE}
The STRIDE code calculates the ideal stability $\delta W$ parameter and inform real-time controllers if the plasma is in danger of surpassing ideal stability limits. This paper will not analyze the accuracy or validity of the calculation as STRIDE has been benchmarked against similar codes in \citet{glasser_riccati_2018} and \citet{glasser_robust_2018}. Adding additional difficulty in benchmarking the real-time versus offline implementation is the differences in equilibria calculated in real-time versus similar faux ``real-time" equilibria calculated post-experiment. While these differences are minor, it has been shown STRIDE is extremely sensitive to equilibria and we would not expect results to align\cite{rothstein_assessing_2025}. 

In the following sections, we begin with discussing the changes to the problem formulation that enables rt-STRIDE to be parallelized in real-time in \autoref{sec:rtSTRIDE} and then show experimental results of the calculation in \autoref{sec:STRIDEexp}. 

\subsection{STRIDE Calculation in real-time}\label{sec:rtSTRIDE}
The key insight that enables a real-time calculation with STRIDE is the reformulation of the ideal stability problem as a Riccati equation rather than a Newcomb equation. This enables us to solve for the state transition matrix (STM), denoted by $\Phi$ with the useful numerical properties that enable starting at some arbitrary initial conditions
\begin{align}
    \mathbf u(0)&=\Phi(0)\mathbf c \\
    \mathbf u(\psi)&=\Phi(\psi)\mathbf c=\Phi(\psi)\mathbf u(0)
\end{align}
the ability to enable us to linearly combine STMs
\begin{align}
    \mathbf u(\psi_2)=\Phi(\psi_2\psi_0)\mathbf u(\psi_0)=\Phi(\psi_2,\psi_1)\Phi(\psi_1,\psi_0)\mathbf u(\psi_0)
\end{align}
and finally the ability to easily switch between forward and backward integration
\begin{align}
    \Phi(\psi_1,\psi_0)=\Phi(\psi_0,\psi_1)^{-1}
\end{align}
For a full description of the problem formulation and different numerical methods, look to \citet{glasser_riccati_2018}, \citet{glasser_ideal_2020}, and \citet{glasser_robust_2018}. 

\begin{figure}
    \centering
    \includegraphics[width=\linewidth]{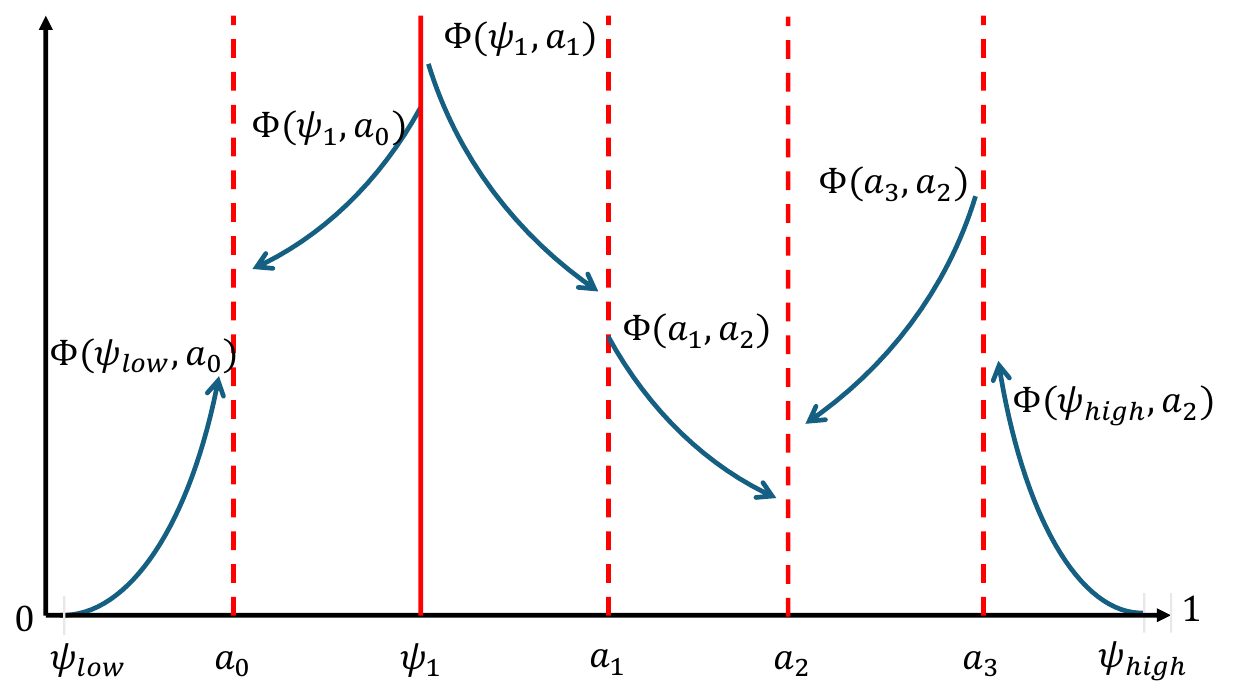}
    \caption{Shooting method of the STM. Each vertical line represents a cut in the integration of the STM. The solid red lines are rational surfaces where the integration of the STM must be split apart and the dashed red lines being artificial interval breaks to create more intervals to divide the calculation further for the \texttt{Workers}. This figure has been adapted from a similar figure in \citet{glasser_robust_2018}.}
    \label{fig:shooting}
\end{figure}

Because of the above properties of the Riccati problem formulation, STRIDE can solve for the STM by a so-called ``shooting" method pictured in \autoref{fig:shooting}. Using this shooting method, it was found the STMs are more numerically stable when calculated moving away from rational surfaces\cite{conlin_mhd_2021}. The real-time relevant aspect of the shooting method is the fact that each STM can be calculated independently and combined at the end, based on the properties listed above. This enables us to use the multi-threading library to let each \worker compute an STM, then the \man can combine all STMs and calculate the final $\delta W$ result from the combined STMs. 

The final problem to address is how to divide the the intervals so when we distribute the calculations to the \workers they take similar times to calculate each interval's STM. The integrators used to calculated $\Phi$ require smaller and more steps close to rational surfaces, so to estimate the required time a measure of Stiffness was assume of the form
\begin{align}
    St=\sum_s\frac{\alpha}{1+\beta|\psi-\psi_s|}
\end{align}
for an interval centered at $\psi$, all singular surfaces $\psi_s$, and tunable parameters $\alpha$ and $\beta$. This stiffness is used to set the distances between some $a_i$ and $a_{i+1}$. $\alpha$ and $\beta$ were selected to as best as possible equally spread the calculation of STMs across \texttt{Workers}. 

\subsection{Experimental results}\label{sec:STRIDEexp}
\begin{figure}
    \centering
    \includegraphics[width=\linewidth]{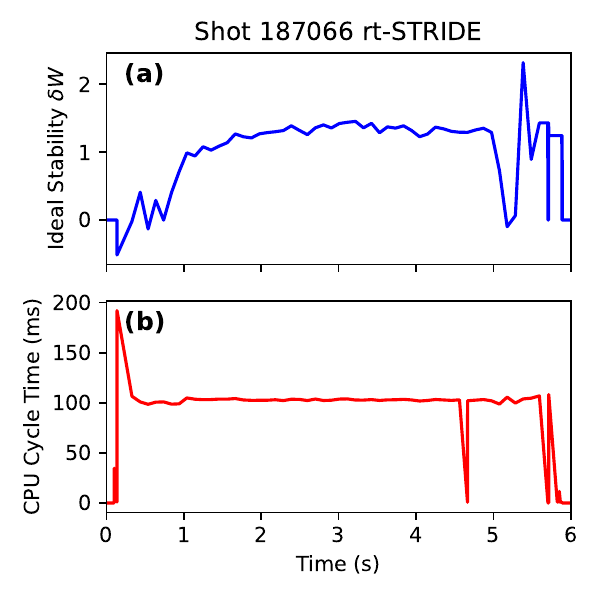}
    \caption{Results from rt-STRIDE calculation in the DIII-D PCS in shot 187066. \textbf{(a)} gives the $\delta W$ stability value over time during the shot and \textbf{(b)} shows the CPU cycle time required for one full rt-STRIDE calculation. }
    \label{fig:rtstride}
\end{figure}

In experiment, rt-STRIDE ran on a $72$ core CPU with approximately $200$ intervals where the STM need be computed. With this approach, rt-STRIDE was consistently calculated in approximately \SI{100}{ms} as seen in \autoref{fig:rtstride}.b with $\delta W$ values given in \autoref{fig:rtstride}.a. There are discrepancies at the beginning and end of the shot and these are attributed to lower quality equilibria causing numerical issues during the plasma current ramp-up and ramp-down. 

Additional testing and benchmarking found that the STM integration time was approximately \SI{20}{ms} of the total \SI{100}{ms} and the majority of the computation time was spent in the serial data pre-processing calculation of equilibrium coordinates to flux coordinates. The largest possible speed-up would likely be  parallelizing this data pre-processing, but further speed=ups could be made with newer hardware with faster processors or more cores.

\section{Conclusions}\label{sec:conclusion}

Two real-time physics codes were efficiently deployed on the DIII-D PCS with a newly implemented, real-time safe multi-threading library. This provided a significant speed-up of these codes to fusion power plant relevant control of ECH power deposition and current drive as well as real-time ideal stability limits. Without the multi-threading capability, these physics codes would not be able to calculate on time-scales relevant for plasma control. 

Future real-time physics-based codes can also leverage this multi-threading library to enable faster real-time calculations of vertical stability\cite{sammuli_avoidance_2021}, other methods for finding ideal wall MHD stability\cite{yang_stability_2025}, or physics-based profile prediction and control\cite{felici_real-time_2011,carpanese_first_2020,morosohk_estimation_2022,morosohk_simultaneous_2025}. Emphasis should be placed on developing physic-based codes that can be parallelized and take advantage of CPU multi-threading or GPU processing. Additionally, machine-learning based models that utilize ensemble averaging\cite{shousha_machine_2023,seo_multimodal_2023} can utilize the individual threads to run independent models and allow the \man to aggregate model results. 

Development and improvements made to real-time codes are crucial for future fusion power plants. Unlike fast machine learning surrogate models, real-time physics codes have robust performance guarantees on day one of operation and can be relied upon without worry of depending on extrapolation techniques. Due to the extra computational costs of physics-based models, additional effort must be devoted to writing efficient PCS code that is deterministic and reliable for error-free, continuous PCS operation. 

\section*{Acknowledgments}

This material is based upon work supported by the U.S. Department of Energy, Office of Science, Office of Fusion Energy Sciences, using the DIII-D National Fusion Facility, a DOE Office of Science user facility, under Award DE-FC02-04ER54698. Additionally, this material is supported by the U.S. Department of Energy, under Award DE-SC0015480. Additionally, this material is supported by the National Science Foundation Graduate Research Fellowship under Grant No. DGE-2039656.

We would like to acknowledge the help from E. Poli and M. Reich in their help of the real-time implementation of the TORBEAM code. We would also like to acknowledge the help from A.S. Glasser and A.H. Glasser for their support of the real-time implementation of the STRIDE code. 

\section*{Disclaimer}

This report was prepared as an account of work sponsored by an agency of the United States Government. Neither the United States Government nor any agency thereof, nor any of their employees, makes any warranty, express or implied, or assumes any legal liability or responsibility for the accuracy, completeness, or usefulness of any information, apparatus, product, or process disclosed, or represents that its use would not infringe privately owned rights. Reference herein to any specific commercial product, process, or service by trade name, trademark, manufacturer, or otherwise does not necessarily constitute or imply its endorsement, recommendation, or favoring by the United States Government or any agency thereof. The views and opinions of authors expressed herein do not necessarily state or reflect those of the United States Government or any agency thereof.

\section*{References}
\bibliographystyle{unsrtnat}
\bibliography{RTSTRIDE.bib}

\end{document}